\documentclass[notitlepage,floats,aps,nofootinbib,preprintnumbers,superscriptaddress,prd,10pt,balancelastpage,twocolumn]{revtex4-1}

\usepackage{amsmath,amssymb,amsfonts}
\usepackage[caption=false]{subfig}
\usepackage{graphicx}
\usepackage{hyperref}
\hypersetup{colorlinks=true, linkcolor=blue, citecolor=red}
\usepackage{color}
\usepackage{url}
\def\beq{\begin{equation}\begin{aligned}}
\def\eeq{\end{aligned}\end{equation}}

\begin{document}
	\title{Optimization of Cost Functions in Absolute Plate Motion Modeling}	
	\author{James Unwin}
        \affiliation{University of Illinois Chicago, 1200 W Harrison St, Chicago, IL 60607, USA}
	\author{Steve Zhang}
        \affiliation{Massachusetts Institute of Technology, 77 Massachusetts Ave, Cambridge, MA 02139, USA}

	\begin{abstract}
We consider the implementation of optimization techniques within the study of tectonic plate motion. Specifically, we examine the optimization underlying optAPM, a leading code for modeling absolute plate motion. We highlight that modifications in the construction of the objective function, composed of individual cost functions, can improve modelling performance. In particular, we propose a simpler and more intuitive formulation of the hotspot cost function. A key part of the new hotspot analysis is the pre-interpolation of hotspot trail data, crucial geological markers for validating absolute plate motion over $\mathcal{O}$(100) Myr timescales. By reducing the propagation of modeling errors, our refined model provides more precise reconstructions of historical plate movements. Our modified hotspot modelling improves the accuracy and reliability of the optAPM outputs. 
	\end{abstract}

	\maketitle


\section{Introduction}
\label{sec:intro}

Obtaining a precise understanding of the evolution of the Earth's tectonic plates back to the earliest eras is challenging. 
The tectonic plates form a dynamic, multicomponent system that is highly complex, especially when one seeks to determine its configuration at timescales of $\mathcal{O}(100)$ Ma.
Computational modelling, drawing on observational data, offers a promising manner to confront such questions. However, the available datasets utilized in plate tectonic studies typically exhibit large spatial and temporal gaps in available data, and measurement errors are difficult to quantify. Moreover, the plate tectonic system as a whole is extremely noisy, characterized by the constant subduction of oceanic plates in conjunction with the formation of new continental crust.
While recent technological advancements, such as satellite geodesy, have improved the accuracy and resolution of current tectonic plate measurements \cite{demets2010geologically, demets1990current}, the reconstruction of historical plate positions is more complicated. Using data such as seafloor spreading records \cite{kusky2013recognition} and isochrons \cite{muller1997digital}, the relative motion of plates can be inferred without direct observation.

Plate motion models can be categorized into two main types: relative plate motion (RPM) models and absolute plate motion (APM) models. 
RPM models describe how plates move with respect to each other, so they constrain plate boundaries but the orientation of whole system is unfixed. In APM models, the plate motions are recorded relative to a fixed reference frame, giving plate velocities relative to the mantle rather than just relative to neighboring plates.
Thus RPM models determine relative rotations up to a net rotation of the whole system, while APM fixes the remaining degree of freedom by choosing a reference frame. 
The strength of APM is that mantle-referenced motions are needed to understand dynamic topography, subduction flux, and net lithosphere rotation. However, the choice of reference frame (e.g.~hotspots, particular plate, no-net-rotation) can introduce nontrivial assumptions, systematic bias, and larger uncertainties, and must be treated with care (cf.~\cite{Clennett,Torsvik,steinberger2008absolute}).

\newpage

In this paper, we investigate the leading absolute plate motion optimization procedure ``optAPM'', developed in Tetley,  \textit{et al.} \cite{tetley2019constraining}. 
OptAPM refines earlier APM models (e.g.~\cite{Gripp,doubrovine2012absolute,Argus,Argus2}) using a comprehensive global inversion incorporating several key constraints: net lithospheric rotation, hotspot motion, and global trench migration. Each constraint is quantified by a cost function, and the process strives to minimize a unified objective function that encapsulates all three. 
The public OptAPM code offers a sophisticated framework for approaching APM through an optimal transport-style approach, which is impressive in its overall scope and implementation. Here, we argue that specific aspects of its implementation are suboptimal and that modifications of the central objective function are justified by both mathematical and heuristic perspectives.
In this study, we approach this process from a more rigorous perspective in an effort to clarify errors and provide a more reliable prediction for the evolution of tectonic plate positions.

This paper is structured as follows:  In Section \ref{sec:constraints}, we explore the various constraints used in optAPM, delineating the cost functions and clarifying the purposes of each. In Section \ref{sec:uncertainties}, we examine the uncertainty inherent in plate motion models by introducing random variations to rotations and observing their propagation over time. In Section \ref{sec:methods}, we identify critical concerns in the optAPM code and present our own optimization methods, contrasting them with the original methods used. In Section \ref{sec:results}, we assess the impact of our methods on model output, with a focus on the hotspot-dependent models. We also review how the three constraints converge into a single objective function, critiquing optAPM's approach at this. We provide concluding remarks in Section \ref{sec:conc}.


\vspace{-5mm}
\section{Constraints on Plate Motion}
\vspace{-2mm}
\label{sec:constraints}

In developing absolute plate motion (APM) models, the common approach is to build upon an existing RPM model. The RPM encompasses all the major tectonic plates and their interactions on Earth's surface. Thus, the foundational RPM model can be utilized to fine-tune the absolute motion of an individual plate in relation to the mantle by optimizing some globally-dependent cost.

\begin{figure}[t!]
    \includegraphics[scale =0.58]{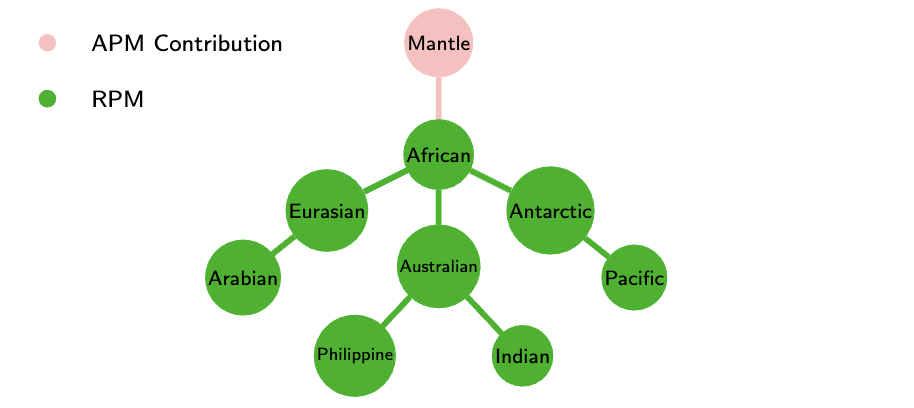}
\vspace*{-4mm}
    \caption{A schematic visualization of a reconstruction tree. Green nodes and edges represent information present in RPM models. The pink node and its corresponding edge represents the additional reference frame provided by the APM model, linking the African plate to the mantle. 
    \label{fig:fig1}}
\vspace{-2mm}
\end{figure}

RPM models, for example  \cite{gurnis2012plate, seton2012global}, represent the dynamics of plate tectonics using a \textit{reconstruction tree}, where nodes symbolize individual plates, identified by unique IDs, and edges define the interactions between neighboring plates. See Figure \ref{fig:fig1} for a visualization of the reconstruction tree. 
The foundational data of RPM models are commonly encapsulated in a rotation model, i.e. a set of finite Euler rotations $(\lambda,\phi,\omega; P, P_{\rm ref}, t)$, where $(\lambda,\phi)$ define the Euler pole, $\omega$ is the rotation angle, and this rotation maps plate $P$ into the reference plate $P_{\rm ref}$ at reconstruction time $t$ (Ma).
This hierarchical structure simplifies the task of deducing the relative motion between any pair of plates to a straightforward traversal of the tree's edges.
RPM models can model any portion of the Earth's surface \cite{gurnis2012plate, seton2012global}. APM models generally build on existing RPM models by establishing a relation between some assigned root plate and an absolute reference frame. 
In the case of optAPM specifically, the designated root-plate is Africa, chosen for its centrality and connection to many major plates, with the mantle serving as the absolute reference frame.

Typically, APM models adopt either the mantle or Earth's spin axis as their absolute reference frame (see e.g.~\cite{Torsvik,Argus}). Models that utilize the spin axis draw on paleomagnetic data to determine the historical alignment of Earth's magnetic poles \cite{Besse,Torsvik-palaeo,mitchell2012supercontinent}. APM that use the mantle as their reference frame, such as optAPM, are primarily informed by the analysis of hotspot trails. To minimize conflicts between disparate data sources, the simplest approach is to restrict modeling to a single constraint. However, beyond just hotspot trails, optAPM incorporates two other constraints, ensuring the model's geodynamic plausibility. This section will detail the purpose of the three constraints and the cost function associated with each.

\vspace{-2mm}
\subsection{Hotspot Trail Misfit}
\label{sec:hotspot}
Hotspot trails form when tectonic plates slide over a mantle upwelling, leaving behind a traceable chain of volcanic formations on the seafloor \cite{TuzoWilson,Morgan1}. These mantle plumes, or hotspots, are considered stationary relative to the mantle for the purposes of plate reconstructions \cite{Gripp,Duncan,maher2015absolute,Steinberger}, due to their minimal velocities when compared to the faster-moving tectonic plates \cite{o2005uncertainties}. As such, when the present-day positions of hotspots are inverted back in time according to plate motion models, the resulting trajectories are expected to align closely with the empirical hotspot trail records. OptAPM uses nine well-documented hotspots spread across the globe: Cobb, Foundation, St.~Helena, Tristan, Reunion, Tasmantid, Samoa, Louisville, and Hawaii. These hotspots have all been studied extensively in prior research \cite{o2005uncertainties, doubrovine2012absolute, knesel2008rapid, koppers2011new, mcdougall1988age, o2013constraints, wessel1998geometric}.

Distance between the predicted and observed hotspot location is computed using the great circle distance between two points on a globe, given by \cite{tetley2019constraining}
 \beq
  d = 2r \sin^{-1} (\sqrt{\varphi}),
\eeq
 with
 \beq
\varphi= \left( \frac{\phi_2 - \phi_1}{2} \right) + \cos(\phi_1) \cos(\phi_2) \sin^2 \left( \frac{\lambda_2 - \lambda_1}{2} \right),
\eeq
where $r$ is Earth's radius, $\phi_1$ and $\phi_2$ are the latitudes of the points, and $\lambda_1$ and $\lambda_2$ are the longitudes. Then the hotspot misfit is calculated via the cost function
\beq
 HS_m = \sum_{j=1}^n \sum_{i=0}^\tau \left(d_{ij}^{(\rm p)}-d_{ij}^{(\rm o)}\right)^{-1} + HS_{gm}\sigma, 
 \label{HSm}
 \eeq
where $d_{ij}^{(\rm p)}$ ($d_{ij}^{(\rm o)}$) is the predicted (observed) great circle distance between time steps $t_i$ and $t_{i-1}$, up to terminal time $\tau$ for hotspot $j\in(1,n)$ (the hotspots sum is implicit in \cite{tetley2019constraining}), and $HS_{gm}\sigma$ is the global hotspot trail misfit standard deviation.  The parameter $HS_m$ seems an unusual construction in the context of optimization theory and thus in Section \ref{sec:methods} we propose an alternative approach.

\vspace{-2mm}
\subsection{Global Trench Migration}
\label{sec:trench}
In a self-consistent APM model, it is expected that subduction zone kinematics are plausible \cite{SCHELLART2008118,Heuret1,Heuret2}. The following three criteria have been shown to be helpful in determining the plausibility of APM models (derived from past models and geodynamic modeling) \cite{flament2017origin,Funiciello,Lallemand,Schellart,SCHELLART2008118}:
{
\parskip=0pt
\begin{itemize}
    \item Maximize the number of retreating trench segments and minimizing the number of advancing trench segments;
    \item Minimize trench migration velocity at the center of wide subduction zones (due to the longer return flow path);
    \item Maximize the number of retreating trench segments near the edges of lateral slabs.
\end{itemize}
}
To compute trench migration in optAPM, plate boundaries are first extracted from the RPM model, and from these, global subduction zones are further extracted and sampled along the boundaries in 1$^\circ$ arc segments. 

At each iteration of the APM inversion, the absolute trench migration vectors orthogonal to each sampled segment are computed. Trench migration misfit is then calculated with the cost function
\beq
 TM_{k} = \frac{1}{n}\sum_{i=1}^n |V_T^{(i)}| + TM_{gT\sigma},
 \eeq
where $|V_T^{(i)}|$ is the trench-normal velocity vector for segment $i$ (summed over segments $1,2,\cdots n$) and $TM_{gT\sigma}$ is the global trench-normal velocity standard deviation. (We have re-expressed the equation in \cite{tetley2019constraining} for clarity).

\vspace{-2mm}
\subsection{Net Lithospheric Rotation}
\label{sec:nlr}
Lastly, the code also incorporates misfit relating to net lithospheric rotation. In past studies, geodynamic flow models provide the lowest estimates of NLR \cite{becker2017superweak}, while models incorporating hotspot data experienced the highest estimates of NLR \cite{o2005uncertainties, torsvik2010plate}. This procedure operates under the assumptions that lower rates of NLR are more likely (although zero NLR is not expected). 
The NLR magnitude is calculated with the cost function (cf.~\cite{torsvik2010plate})
\beq
  \omega_{\rm net} = \frac{3}{8\pi r^4} \sum_{i} \int (\omega_i \times R) \times R ~{\rm d}S_i,\eeq
where $\omega_{\rm net}$ is the calculated NLR rate in degrees per million years, $r$ is the radius of the Earth, $R$ is the plate rotation velocity vector, ${\rm d}S$ is the area element integrated over the sphere, and $\omega_i$ is the plate angular velocity vector for a given plate $i$. The base optAPM  \cite{tetley2019constraining} strongly favours models that keep $\omega_{\rm net}$ small, imposing a heavily weighted linear cost for increasing $\omega_{\rm net}$.

\subsection{Objective Function}
\label{sec:objective}
For each iteration of the optimization process, the model optimizes absolute motion at 5 Myr intervals. In particular, it optimizes for the Euler rotation of Africa relative to the mantle, which minimizes a cost determined by the objective function
\beq
\label{1}
 J = \frac{HS_m}{\sigma_1} + \frac{TM_k}{\sigma_2} + \frac{\omega_{\rm net}}{\sigma_3},\eeq
where $\sigma_1$, $\sigma_2$, and $\sigma_3$ are the relative weighings of each constraint. The determination of the Euler rotation involves the propagation of various Euler poles across the globe. At each of these poles, the model calculates the optimal rotation angle, effectively setting a seed for the iterative search that follows for a local minimum of the cost function. To enhance the model's resolution, additional seed poles can be introduced. More details on this optimization technique are given in Section \ref{sec:methods}.

In formulating the objective function of eq.~(\ref{1}),  one must combine the individual cost functions. Notably, the formulation of optAPM outlined in \cite{tetley2019constraining} appears to combine the individual cost functions via a seemingly {\em ad hoc} weighted sum with
\beq
\sigma_1=8, \qquad
\sigma_2=1, \qquad
\sigma_3=\frac{1}{125}.
 \eeq
Note that smaller values imply higher weighting.
The rationale behind these specific scaling values is unjustified, which raises some concerns. A potential approach to systematically fix $\sigma_i$ is via a chi-square-informed weighting, but implementing this is beyond the scope of the present work. In what follows, we either adopt the above values (to allow comparison with \cite{tetley2019constraining}) or alternatively, we take $\sigma_i=1$ for $i\in(1,3)$ with $\sigma_{j\neq i}\rightarrow\infty$  in order to isolate individual constraints for examination.


\vspace{-2mm}
\section{Model Uncertainties}
\label{sec:uncertainties}
The task of constraining absolute plate motions is challenging, largely due to the discrete nature of geological data (both spatially and temporally), leading to many inaccuracies and uncertainties in the time-periods in between known data. In particular, the \textit{world uncertainty}, or the percentage of the Earth's surface that has subducted since a given time, reaches as high as 60\% by 140 million years ago (Ma) \cite{torsvik2010plate, muller2016ocean}. As such, available hotspot data, predominantly existing on oceanic plates, also decreases significantly as we look further back in time. Methods for retrieving such subducted plates exist, such as the use of seismic tomography to locate large low-shear-velocity provinces \cite{van2010towards,Garnero}. Despite this, the data related to these geological anomalies remains imprecise, making direct reconstructions challenging \cite{Hellinger}.

The exact positions and velocities of present-day hotspots, key factors in plate motion models, are still debated \cite{wang2017bounds,Tarduno}. Furthermore, hotspots are not entirely stationary relative to the mantle, and the velocity of rising mantle plumes is variable \cite{Steinberger,doubrovine2012absolute}. Hotspots are also disproportionately concentrated on the Pacific plate \cite{Clouard,Gripp}, potentially introducing the risk of sampling bias. In the analysis of \cite{tetley2019constraining}, four out of the nine hotspot trails used to constrain the data lie on the Pacific plate. To investigate potential discrepancies between hotspot trails, which deviate from each other at rates of up to 50 mm/yr \cite{wang2017bounds,o2005uncertainties}, we generated models incorporating every subset of seven hotspots of the nine total hotspots.

\begin{figure}[t!]
    \centering
    \includegraphics[scale = 0.54, trim = 2cm .5cm 2cm 0cm]{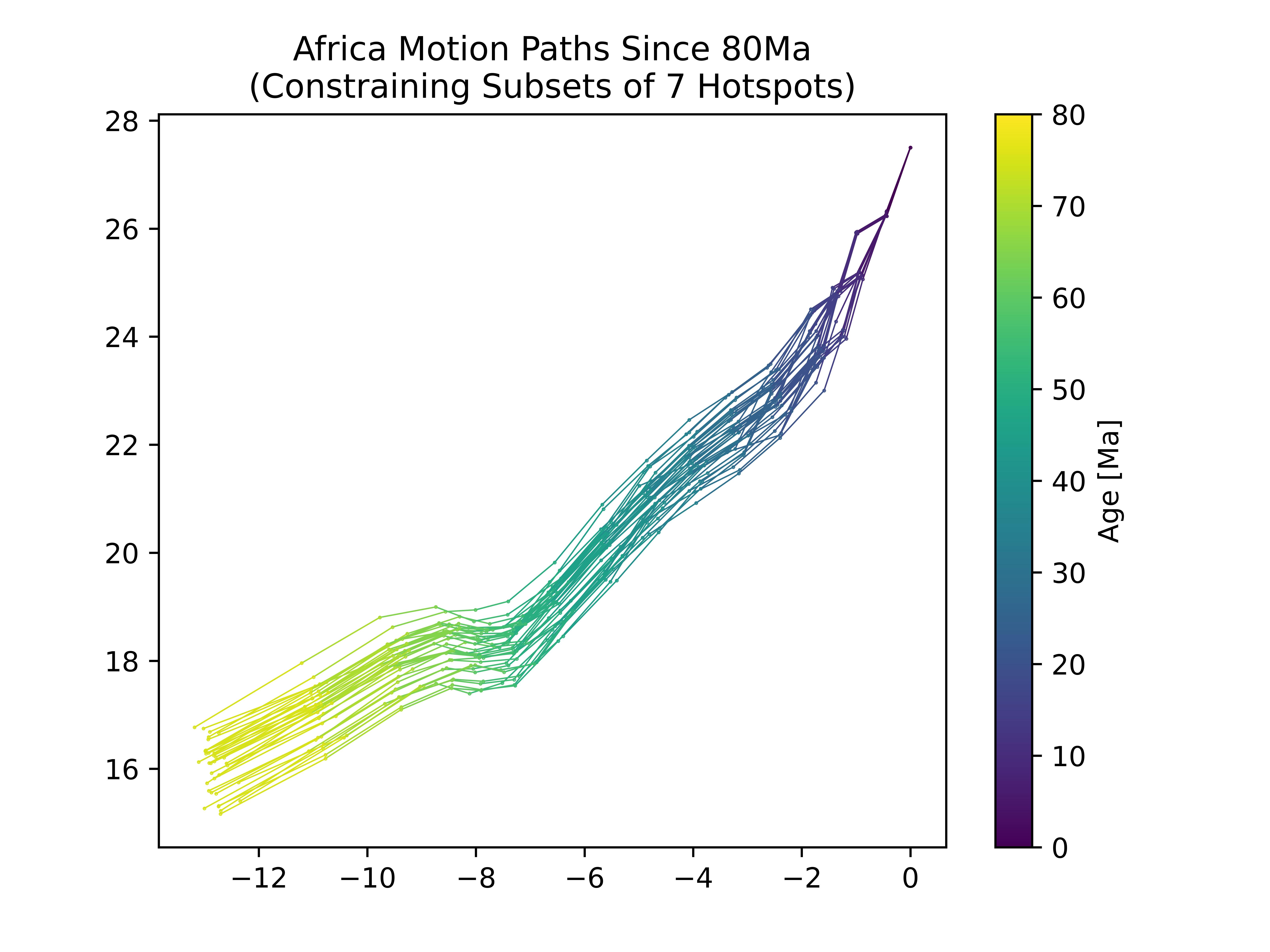}
\vspace{-3mm}
    \caption{Predicted motion paths when the point ($0^\circ$N, $27.5^\circ$E) in the African plate is reconstructed back to 80 Ma for 36 different models, each incorporating a unique subset of 7 hotspots from the 9 total hotspots utilized in optAPM. Color represents age in millions of years, and individual line segments between adjacent points represent absolute motion in a 5 Myr interval period. The aim of the plot is to highlight the compounding effect of small deviations.
    \label{fig:hotspot_subsets}}
\vspace{-4mm}
\end{figure}

Figure \ref{fig:hotspot_subsets} illustrates the reconstructed motion paths derived from 36 distinct models, each constrained by a unique subset of seven out of the nine hotspots employed by optAPM. The incremental progressions of these paths are largely consistent across the different models, a similarity attributed to the governing constraints of net lithospheric rotation (NLR) and trench migration (TM). 
Nevertheless, there is a noticeable divergence among the models, with a dispersion of up to  $\sim3^\circ$ in longitude, which can be traced back to variations in hotspot trail data. This divergence underscores the role of hotspots in fine-tuning the model outputs, while also emphasizing that the overarching trajectory is shaped by NLR and TM constraints.
Additionally, the large amount of data informing a single plate motion model complicates the quantification of model uncertainty using standard metrics \cite{Kirkwood}. Since model output relies on its preceding results, errors inherently compound and scale up as models date further back \cite{Jurdy}.

\begin{figure}[t!]
    \centering
    \includegraphics[scale = 0.42, trim = 0cm 0cm 0cm 0cm]{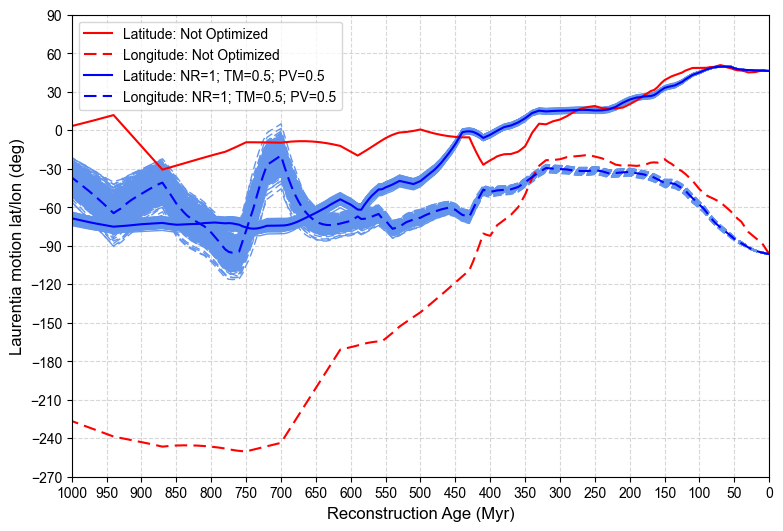}
  \vspace{-1mm}
    \caption{Reconstruction path of a reference point in Laurentia over the past 1000 Ma, derived from models with introduced variability. The intermediate rotation angle $\omega$ is adjusted at 5 Myr intervals using the function $\omega (1+0.1z)$, where $z$ is a random variable derived from a Gaussian distribution centered at $0$ with standard deviation $1$. The red lines depict the latitudinal and longitudinal paths from the unoptimized model, while the dark blue lines show the same for the optimized model. The shaded blue areas represent the variability among 1000 perturbed model.
    \label{fig:laurentia}}
\vspace{-3mm}
\end{figure}

As a warm-up demonstration, we introduce controlled variability into the plate model of \cite{merdith2017full}. We perturb the rotation angle of Laurentia relative to the mantle to be  $ \omega(1+0.1z),$
where $\omega$ represents the original rotation angle, and $z$ is a variable derived from a Gaussian distribution centered at $0$ with unit standard deviation. This process of error introduction was repeated across each 5-million-year interval, thereby simulating the accumulation of error back to 1000 Ma. (Typically elsewhere in this paper, we consider shorter time periods $\sim80$ Ma).
Figure \ref{fig:laurentia} illustrates the variations in Laurentia's latitudinal and longitudinal movements over 1000 simulations. The unoptimized rotation model's trajectories (from the RPM \cite{merdith2017full}) are shown as red lines, while the model's trajectories optimized via optAPM \cite{tetley2019constraining} are shown in blue.  As expected, Laurentia experiences lower rates of rotation in the optimized model. The paler blue shaded areas surrounding the optimized paths indicate the range of plate motions across the 1000 perturbed (via $z$) models. Notably, at 1000 Ma, the spread of possible positions for Laurentia spans approximately 35 degrees in latitude and about 10 degrees in longitude. The above is intended to illustrate that compounding errors can lead to large deviations over long periods (as discussed in e.g.~\cite{Kirkwood,Jurdy}).


\section{Methods and Model Setup}
\label{sec:methods}
\begin{figure*}[t!]
    \vspace*{5mm}
    \includegraphics[scale = 0.21]{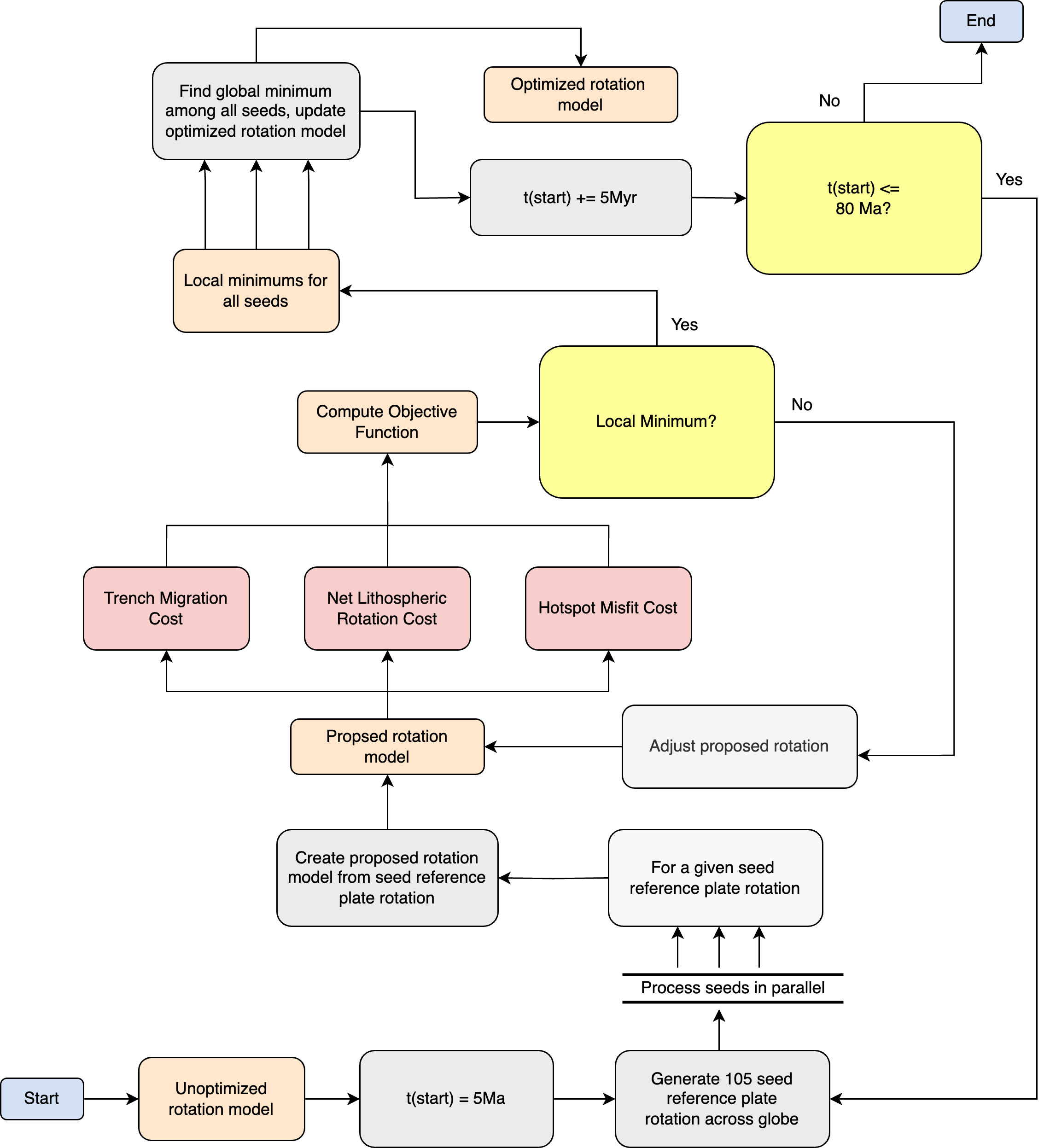}
    \vspace{15mm}
    \caption{Workflow diagram illustrating the full optimization process (emulating the diagramatic style in \cite{Muller}). The code optimizes absolute plate motion at 5-million-year (5Myr) intervals, dating back to 80 Ma. The African plate is used as the reference throughout the procedure, chosen for its central position and close relationship with other major plates. At each interval, the code propagates a collection of 105 seed Euler rotations, uniformly distributed within a 60$^\circ$ radius of the preceding optimal Euler pole. For each seed rotation, the algorithm creates a temporary rotation model. This model is then used to calculate the value of the objective function, guiding the algorithm through iterative adjustments toward a local minimum. The process then determines a global minimum from these local minima, incrementally refining the plate motion model. 
    \label{fig:flowchart}}
    \vspace*{5mm}
\end{figure*}

For our reconstructions, we adapted the publicly available optAPM code \cite{tetley2019constraining}, implementing substantial enhancements to both the optimization process and the cost function definitions. Our chosen RPM model from \cite{merdith2017full} offers a comprehensive global reconstruction dating back to 1000 Ma (see also \cite{GPlates,Muller,Boyden}). In addition to describing relative motion between plates, this model captures topological changes over time, providing a nuanced picture of net lithospheric rotation and enabling more precise calculations.
Our code optimized absolute plate motion at 5 Myr increments. Our reconstructions were limited to the past 80 million years, reflecting the consensus that hotspot data beyond this point lacks sufficient reliability (see e.g.~\cite{Shimin,steinberger2008absolute,koppers2011new}). At each interval, the code computes the Euler rotation of the African plate relative to the mantle, which minimizes the objective function, and updates the APM model. A detailed visual representation of this modeling process is available in the flowchart depicted in Figure \ref{fig:flowchart}.

To identify local minima from seed poles, our method utilized the COBYLA algorithm \cite{COBYLA} (Constrained Optimization BY Linear Approximations) as integrated within the NLopt optimization library \cite{johnson2021nlopt}. For each interval, our method propagated a collection of 105 seed Euler rotations, uniformly distributed within a 60$^\circ$ radius of the preceding optimal Euler pole. 
Iteratively, COBYLA systematically searched for a local minimum of the objective function value, and our method then extracted the optimized minimum. COBYLA is particularly effective in complex, constraint-rich environments and operates independently of gradient information.

\clearpage

\noindent
We imposed a cap of 1000 iterations on runs that were especially computationally demanding, such as those involving isolated trench migration. We verified that generically 
COBYLA to reached full convergence to a local minimum within this limit and that this cap does not materially impact our results.

\begin{figure}[t!]
    \centering
\hspace*{-3mm}  \includegraphics[scale = 0.42, trim = 2cm 1.5cm 2cm 0cm]{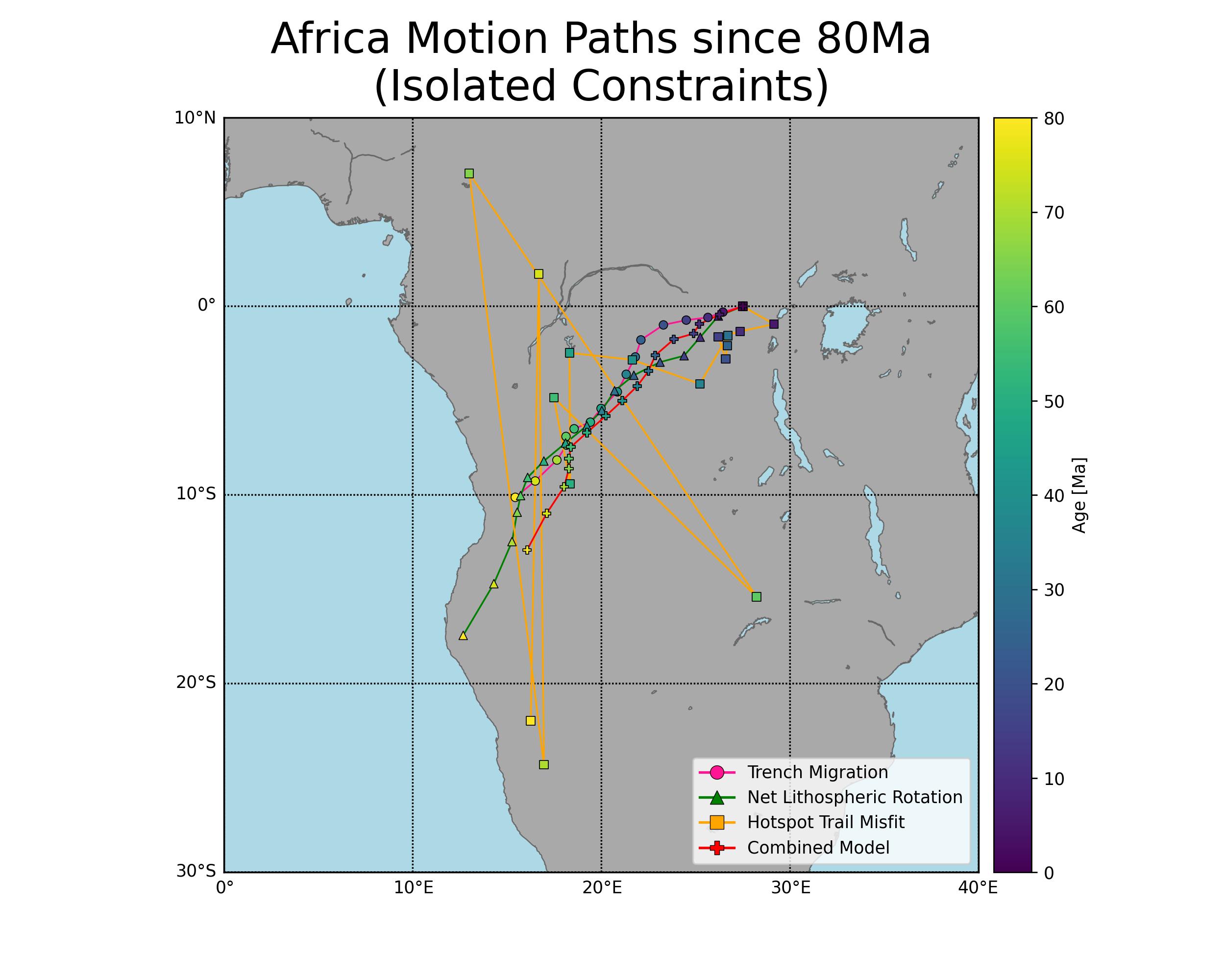}
\vspace{-5mm}
    \caption{Predicted motion paths when the point at ($0^\circ$N, $27.5^\circ$E) in the African plate is reconstructed back to 80 Ma according to three different models, each isolating a different model constraint: trench migration (pink circles), net lithospheric rotation (green triangles), and hotspot trail misfit (orange squares). Additionally, the red plus signs indicate the integrated model that considers all three constraints simultaneously. The models operate on 5 Myr intervals, with varying colors denoting the age of the reconstructed position.
    \label{fig:isolated_constraints}}
\vspace{-2mm}
\end{figure}

\vspace{-2mm}
\subsection{Isolating Constraints}
\label{sec:isolating}
To enhance the robustness of the optimization procedure, we conducted an examination of the models resulting from isolating individual constraints within the code. We consider the objective function (of eq.~(\ref{1}))
\beq \notag
 J = \frac{HS_m}{\sigma_1} + \frac{TM_k}{\sigma_2} + \frac{\omega_{\rm net}}{\sigma_3},
 \eeq
and set two of the weights ($\sigma_1$, $\sigma_2$, and $\sigma_3$) to infinity, while fixing the third at $1$. We will later replace $HS_m$ with a new hotspot cost function $\overline{HS_m}$. 

Figure \ref{fig:isolated_constraints} shows the resulting motion paths for each of the three cases, as well as a model combining all three, to reconstruct a point ($0^\circ$N, $27.5^\circ$E) in the African plate. Notably, significant issues with the hotspot cost function became apparent when this function was singled out during optimization. Specifically, the hotspot-only model predicted an average velocity of 22.1 cm/yr, in contrast to the average velocities of  2.8 cm/yr and 2.2 cm/yr from the models, isolating trench migration and net lithospheric rotation, respectively. We address these discrepancies in the next subsection.

 \vspace{-2mm}

\subsection{Changes to Cost Functions}
\label{sec:changestocost}
 \vspace{-2mm}

We first look to refine the optAPM model by modifying the cost functions associated with hotspot data and trench migration. The original cost functions employed standard deviation as a measure for these constraints, a choice that was not sufficiently justified within the existing documentation. To achieve a more straightforward and transparent set of criteria, we utilize the average for both constraints, as detailed below. 

The original hotspot cost function followed the steps; specifying a specific hotspot $j$:
\begin{enumerate}
    \item Create rotation model at reconstruction time $t_i$.
    \item Search for all hotspot trail data within a 10~Myr interval succeeding the reconstruction time.
    \item Interpolate the model output between  $t_i$ and $t_{i+1}$, define a smooth function $\Psi(x,y,t)$ for latitude $x$, longitude~$y$, and time $t$.
    \item For all hotspot data points $t_{\rm hs}$, at the \textit{hotspot data time}, compute difference between:
    \begin{itemize}
        \item Raw hotspot trail data ($d_{ij}^{(\rm o)})$) at $t_{\rm hs}$, and
        \item Interpolated model output ($d_{ij}^{(\Psi)}$) at $t_{\rm hs}$.
    \end{itemize}
    \item Summing over $j$, minimize $HS_m$ (cf.~eq.~(\ref{HSm})).
\end{enumerate}

In our revised model, we implemented a new hotspot cost function via the following steps:
\begin{enumerate}
    \item Create rotation model at reconstruction time $t_i$.
    \item Interpolate all hotspot data, defining a smooth function $\Phi(x,y,t)$.
    \item For all hotspot trails, at the \textit{reconstruction time}, compute distance between:
    \begin{itemize}
        \item Interpolated hotspot trail  ($d_{ij}^{(\Phi)}$) at $t_{\rm i}$, and
        \item Raw model output ($d_{ij}^{(\rm p)}$)  at $t_{\rm i}$.
    \end{itemize}
    \item Minimize $\overline{HS_m}$ = $\frac{1}{n}\sum_{j}^n\sum_{i}^\tau (d_{ij}^{(\rm p)}-d_{ij}^{(\Phi)})$.
\end{enumerate}
While the original model seeks to optimize the interpolation points interspersed among the model's outputs, our modified model focuses directly on optimizing the model outputs themselves.   The aim of this modification is to mitigate the accumulation of errors over time, therefore partially mitigating the issues exhibited in Figure \ref{fig:hotspot_subsets}. 

\begin{figure*}[t]
  \centering
  \includegraphics[width=0.83\textwidth,trim=0cm 0cm 0cm 1cm,clip]{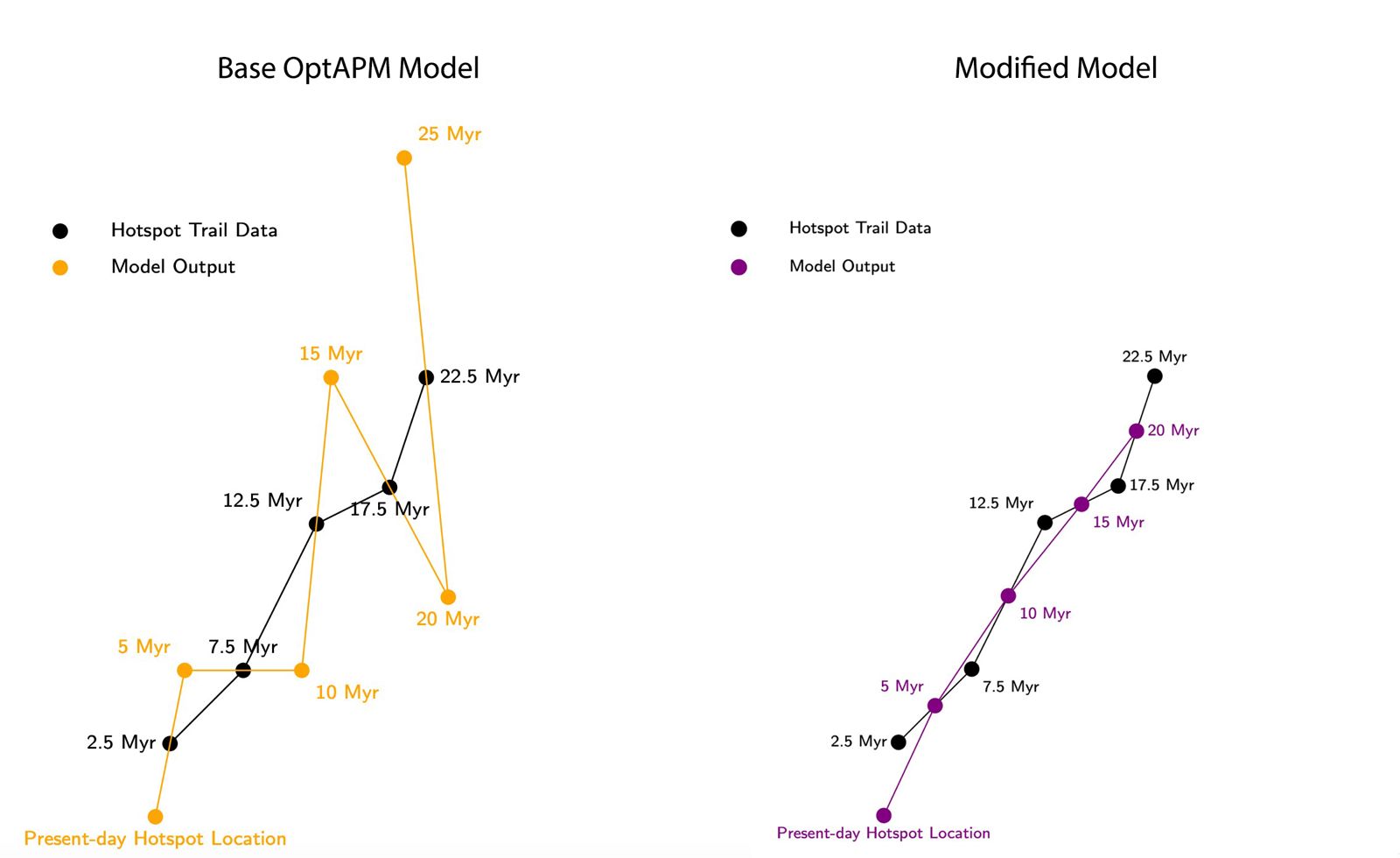}
  \caption{A comparative illustration of model performance before and after the implementation of the new hotspot cost function, focusing on a single hotspot trail for clarity. The black lines represent the actual hotspot trail data. On the left, the orange trajectory shows the model output using the original cost function, which minimizes distances at times corresponding to raw hotspot trail data. On the right, the purple trajectory depicts the model output using the revised cost function, which minimizes distances precisely at the reconstruction times. In this special case, the model output of the revised hotspot cost function perfectly matches the interpolated hotspot trail data.}
  \label{fig:interpolation}
\vspace{-5mm}
\end{figure*}

Figure \ref{fig:interpolation} shows an outlier case in which all hotspot trail data falls between multiples of 5 Myrs. By streamlining the model to consider only a single hotspot trail, it becomes clear that the revised model, adhering to the new cost function, maintains a trajectory that is more consistent with the hotspot trail data since the original model fluctuates significantly around the trail data.

\newpage

\onecolumngrid
\twocolumngrid

\section{Results and Analysis}
\label{sec:results}

\begin{figure*}[p]
  \centering

  \subfloat[ 
   Comparison between raw hotspot trail data and interpolated hotspot trail data in two well-studied hotspots: Reunion and Louisville. Circles represent the raw hotspot trail data and triangles represent the interpolated hotspot trail data. Color represents age in Ma.
  ]{%
    \begin{minipage}[t]{\textwidth}\vspace{0pt}
      \centering
      \begin{minipage}[t]{0.49\textwidth}\vspace{0pt}
        \centering
        \includegraphics[
          width=\linewidth,
          height=0.30\textheight,
          keepaspectratio,
          trim=3cm 1cm 3cm 0cm,clip
        ]{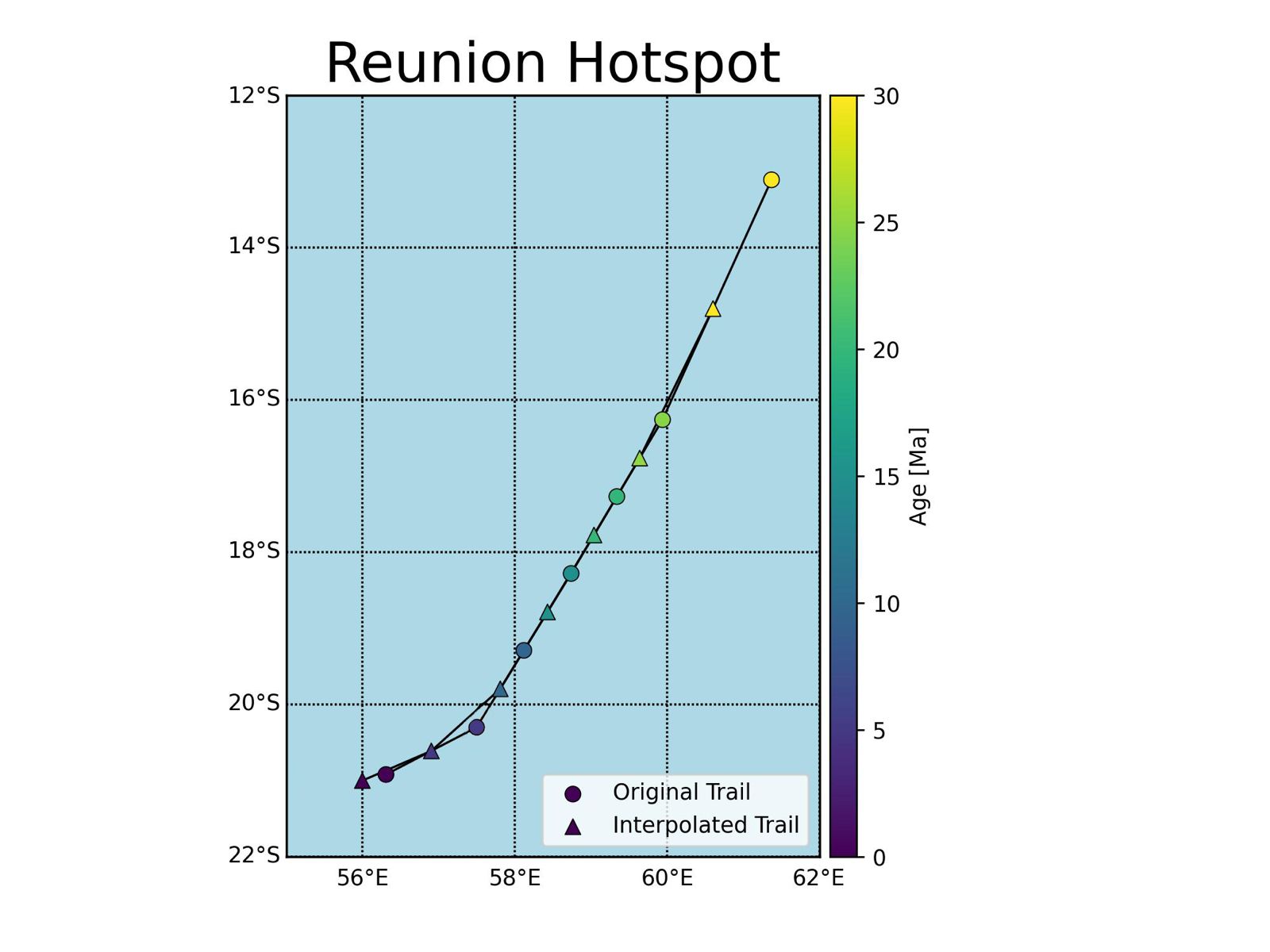}
      \end{minipage}
      \hfill
      \begin{minipage}[t]{0.49\textwidth}\vspace{0pt}
        \centering
        \hspace*{-1cm}
        \includegraphics[
          width=1.1\linewidth,
          height=0.30\textheight,
          keepaspectratio,
          trim=1cm 1cm 2cm 0cm,clip
        ]{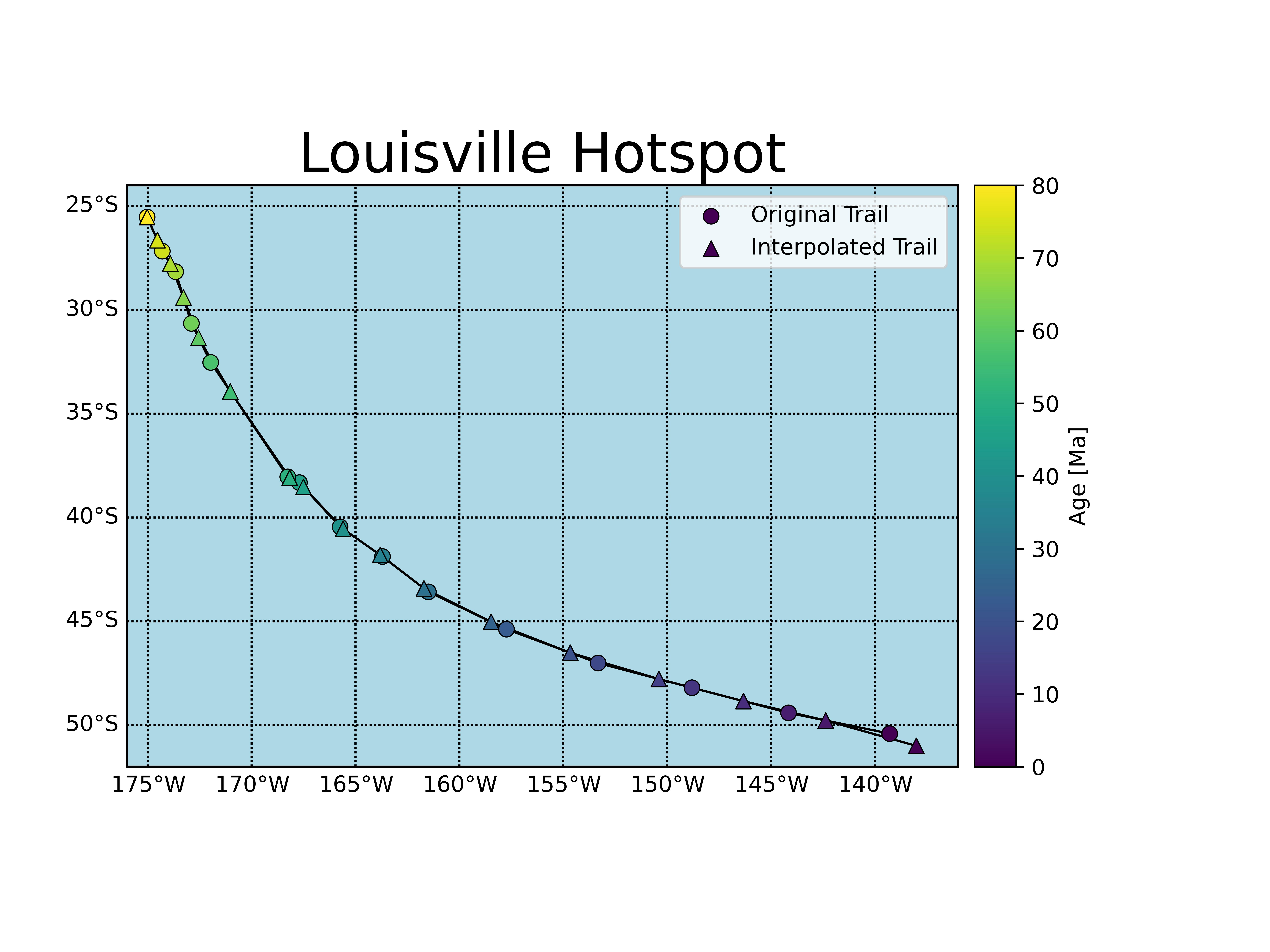}
      \end{minipage}
   \vspace{4mm}
    \end{minipage}%
    \label{fig:hotspots}
  }

  \vspace{8pt}

  \subfloat[
      Predicted motion paths, comparing the base OptAPM hotspot cost function (marked by squares and orange path) and our modified hotspot cost function (marked by ``X'' and purple path). Other aspects follow the same data-generation process as Fig.~\ref{fig:isolated_constraints}.
  ]{%
    \begin{minipage}[t]{0.49\textwidth}\vspace{0pt}
\vspace{15mm}
      \centering
      \includegraphics[
        width=\linewidth,
        height=0.44\textheight,
        keepaspectratio,
        trim=2cm 1.5cm 2cm 0cm,clip
      ]{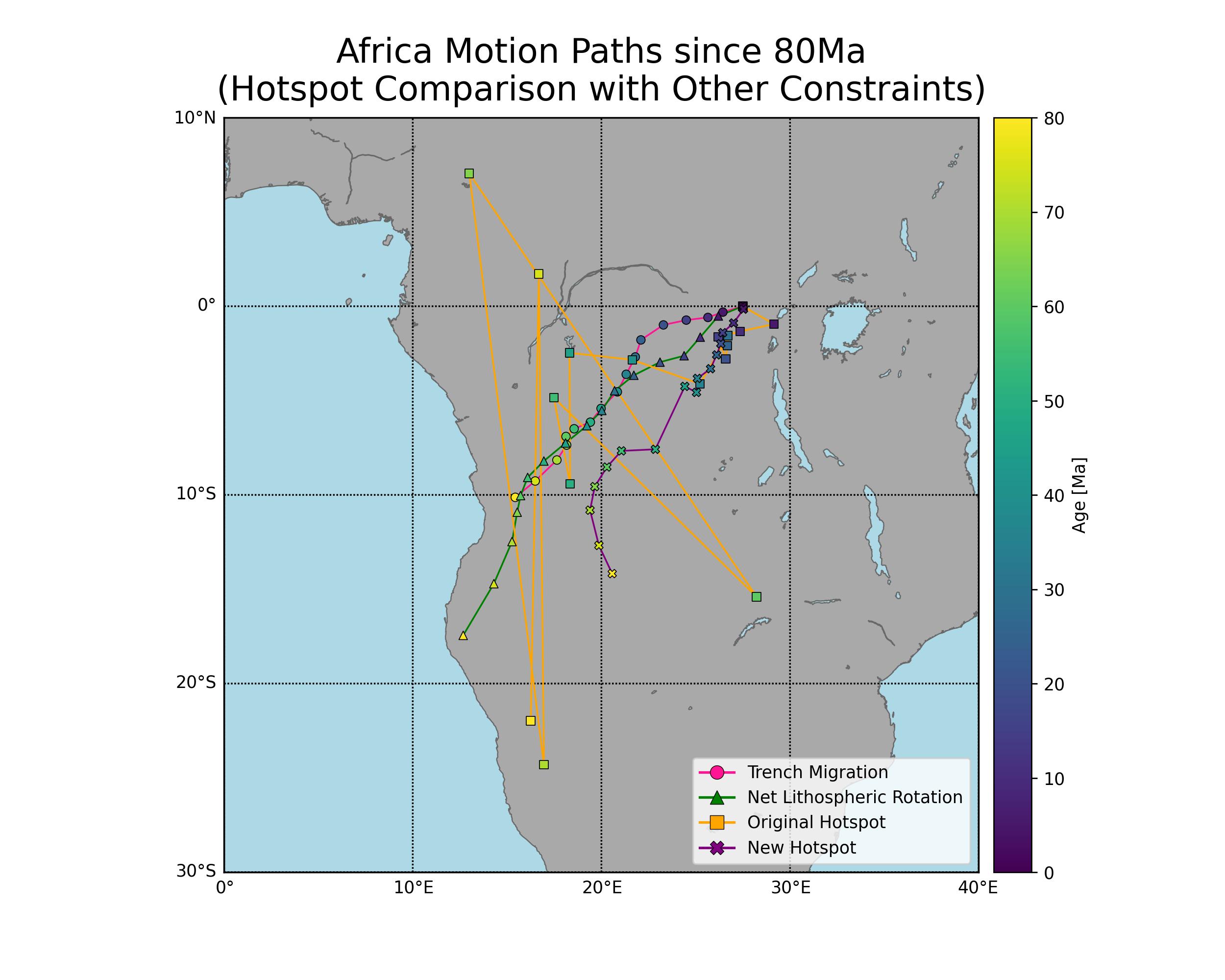}
\vspace{3mm}
    \end{minipage}%
    \label{fig:with_others}
  }
  \hfill
  \subfloat[
 A comparison of the net lithospheric rotation and median trench migration as predicted by the hotspot models of the base optAPM code \cite{tetley2019constraining} and the modified approach presented here. Orange lines  (old) show outcomes from the model using the original hotspot cost function, while the purple lines  (new) show results from our methodology. The shaded regions around median trench migration indicate the range of trench migration velocities at given times.
    \label{fig:otherconstraints}
  ]{%
    \begin{minipage}[t]{0.49\textwidth}\vspace{0pt}
      \centering
      \includegraphics[
        width=\linewidth,
        height=0.205\textheight,
        keepaspectratio,
        trim=1cm .5cm 1cm 0cm,clip
      ]{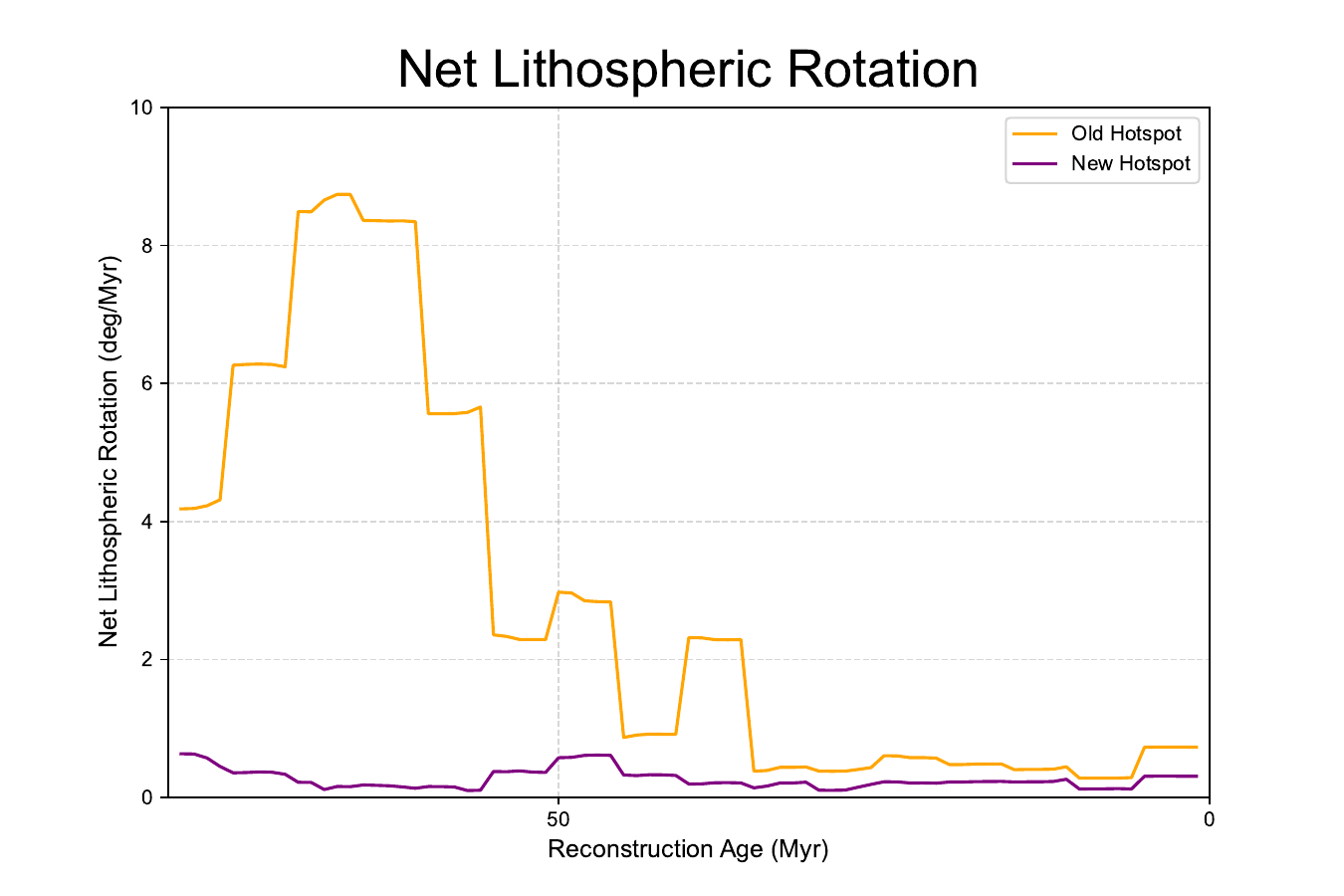}

      \vspace{6pt}

      \includegraphics[
        width=\linewidth,
        height=0.205\textheight,
        keepaspectratio,
        trim=1cm .5cm 1cm 0cm,clip
      ]{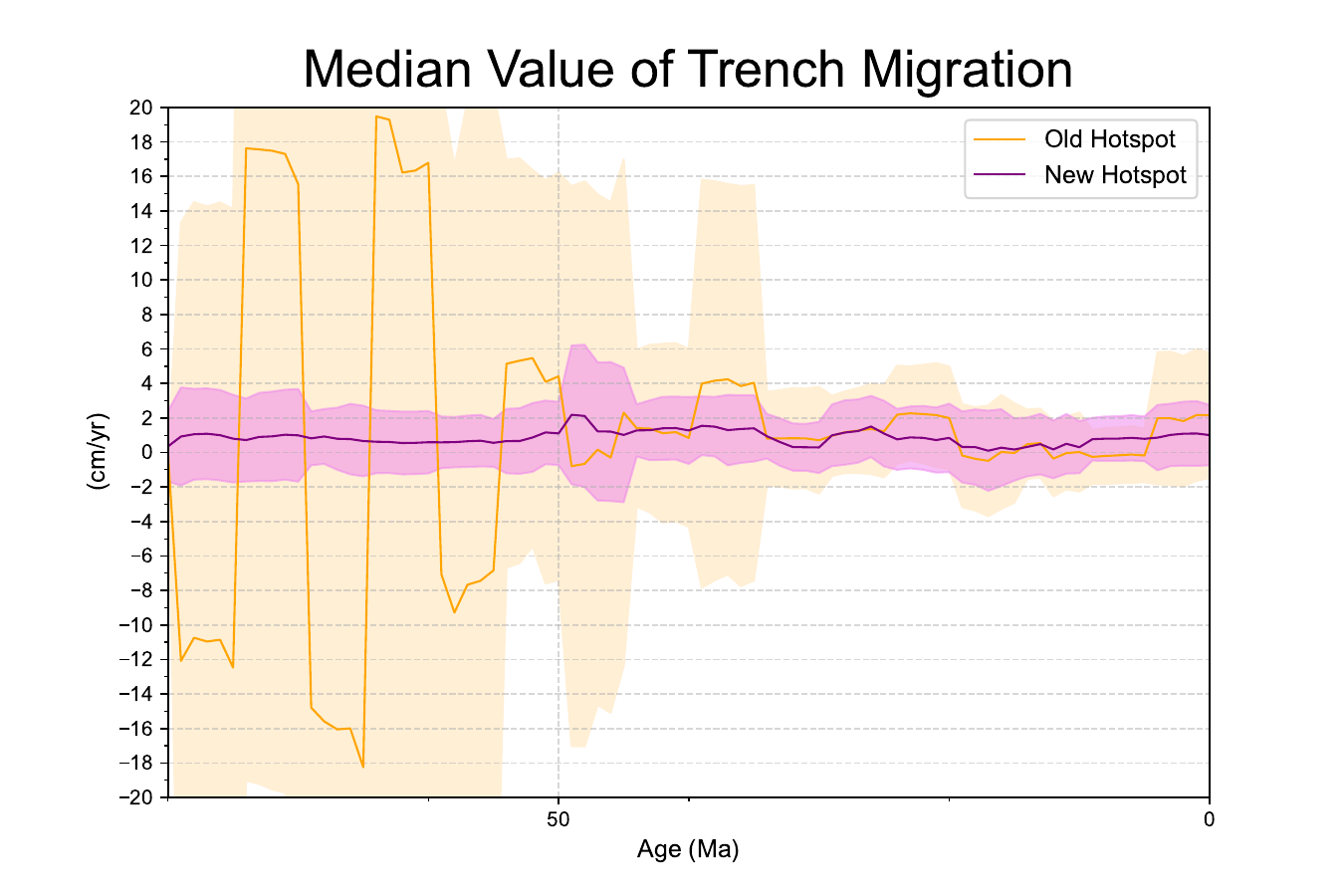}
    \end{minipage}%
    \label{fig:otherconstraints}
  }

  \caption{Hotspot trail comparison and associated diagnostic constraints used in the APM optimization.}
  \label{fig:minipage}
\end{figure*}

 To assess the impact of our modifications, we reintegrated the three cost functions and generated a revised model. Subsequently, we conducted a comparative analysis between this our new model, now includes the refined hotspot cost function, and the base optAPM code  \cite{tetley2019constraining}. 
Following the procedure outlined in Section \ref{sec:changestocost}, we begin our analysis by interpolating all hotspot data to 5-million-year increments. After performing this interpolation, we derive our model. 
Since tectonic plates typically traverse linear paths and hotspots are approximately fixed in relation to the mantle, the resulting hotspot data exhibited a relative smoothness. The interpolated trails yield trajectories that remain consistent with the raw hotspot trails, as depicted in Figure \ref{fig:hotspots}. 
While changes in hotspot trail data are minimal, the effects on model output can be significant.

We run our modified version of optAPM; for all runs, we use the terminal date 80 Ma. 
Figure \ref{fig:with_others} presents a similar analysis to Figure \ref{fig:isolated_constraints}, where we apply each constraint in isolation. However, this time, we employ our improved hotspot modeling and compare it to the original model of \cite{tetley2019constraining}.
As in Figure \ref{fig:isolated_constraints}, we reconstruct the point ($0^\circ$N, $27.5^\circ$E) in our diagrams.
These changes reflect the trends seen in Figure \ref{fig:interpolation}, where the modified model demonstrates a more consistent and linear progression compared to \cite{tetley2019constraining}.

The original model estimated the African plate's average absolute velocity to be 22.1 cm/year, with an average angular variation of 57.5 degrees over 5 Myr periods. In contrast, the revised model reported a significantly reduced average plate velocity of approximately 2.6 cm/year and an angular variation of 17.0 degrees, indicating a substantial improvement in consistency.

The refined hotspot-only model exhibited a significantly improved alignment with other geological constraints, cf.~Figure \ref{fig:otherconstraints}. 
The new model predicted an average trench migration velocity of 0.87 cm/year, a marked reduction from the  \cite{tetley2019constraining} estimate of 5.42 cm/year. Moreover, all median trench migration values were positive and less than 2 cm/year, aligning with geodynamic models that suggest a low but positive rate of trench migration. The standard deviation for trench migration in the modified model was also significantly reduced to 0.40 cm/year, in contrast to the earlier 8.23 cm/year, indicating a more stable and consistent modeling outcome.

In terms of NLR, the modified model presented a rate of 0.26$^\circ$/Myr, much lower than the 2.78$^\circ$/Myr in \cite{tetley2019constraining}. This rate is in agreement with historical hotspot-based models, which typically yield NLR values between 0.10 and 0.50$^\circ$/Myr. Nonetheless, it is noted that the NLR arising in our model marginally exceeds the larger upper bounds from geodynamic studies which constrain NLR $\lesssim0.2-0.4^\circ$/Myr \cite{Conrad,Becker,Muller,Gripp}.


\clearpage

\begin{figure}[t!]
    \centering
   \includegraphics[scale = 0.4, trim = 2cm 2cm 2cm 0cm]{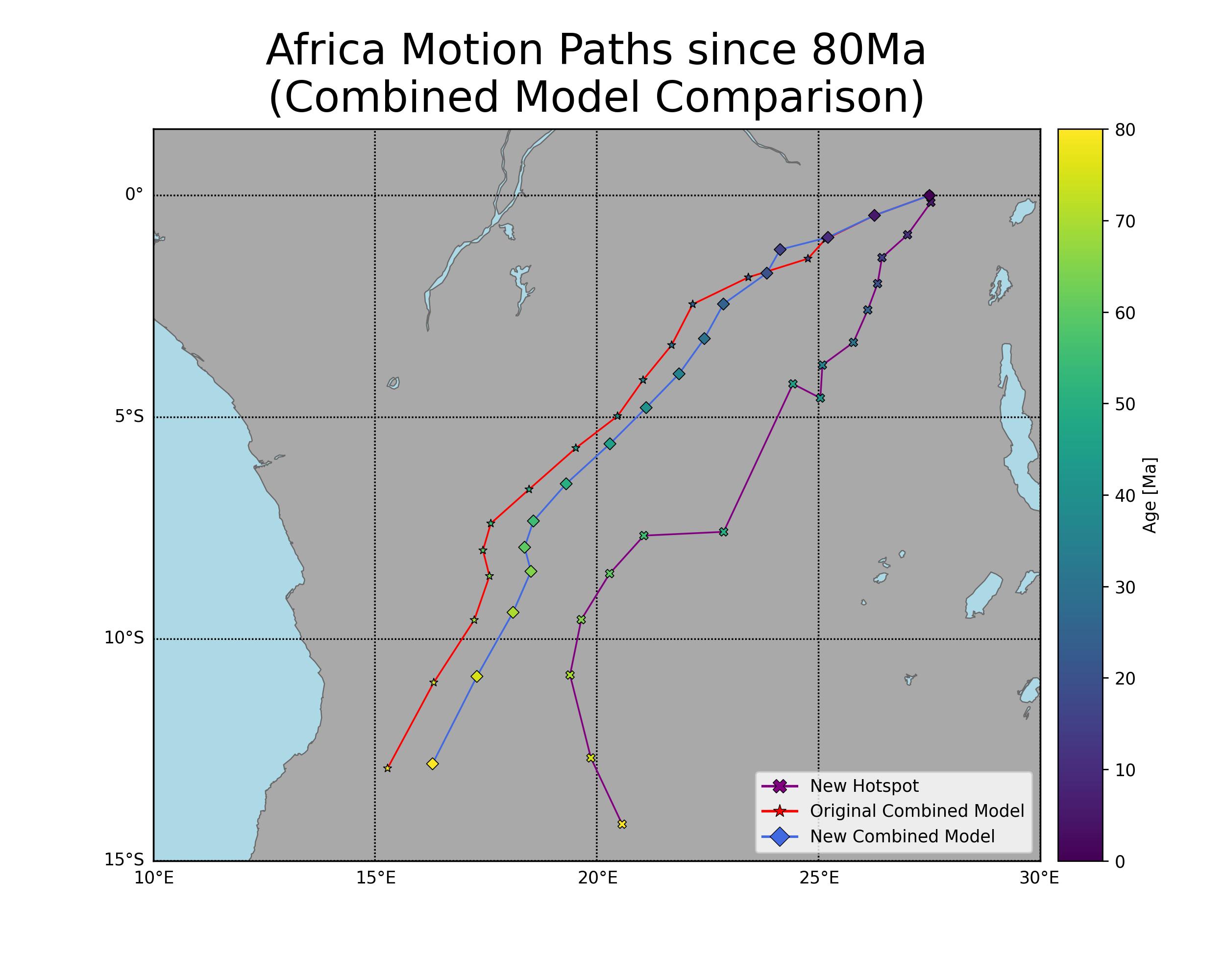}
    \caption{
   Predicted motion (similar to Figs.~\ref{fig:isolated_constraints} \& \ref{fig:with_others}), comparing the modified hotspot cost function (marked by ``X'' and purple path) to the
   base OptAPM objective function (stars and red path) and our modified objective function which uses the new hotspot cost function (diamonds and blue path). 
       \label{fig:combinedcomparison}}
\vspace{-4mm}
\end{figure}

Finally, in Figure \ref{fig:combinedcomparison} we illustrate the effectiveness of the revised hotspot cost function within a comprehensive plate motion model that integrated all three constraints. The model incorporating the new cost function, indicated by blue diamonds, showed a reduced average deviation of 315 km from the expected path: a notable improvement over the original model's deviation of 379 km, denoted by red stars. This suggests a refinement in the model's ability to predict and minimize errors across various hotspot tracks, likely due to the more streamlined optimization process.

\newpage
\section*{Summary and Concluding Remarks}
\label{sec:conc}
Plate motion modelling aims to map the positions of the tectonic plates millions of years ago. This offers insights into Earth's geological evolution at local and global scales. The computational framework of optAPM implements a sophisticated optimization approach and is an impressive feat of coding. Here, we have endeavored to demonstrate that additional fine-tuning can further enhance the reliability and consistency of APM modeling.

In this work, we began with an analysis of the sensitivity of plate motion models, this was performed by observing the effects of introducing statistical noise to the rotation angle at model intervals. We then isolated individual constraints within the objective function to gauge the relative impact of each. This approach highlighted major discrepancies among the various outputs. After identifying pitfalls in the cost functions, we outlined a refinement of the original optAPM laid out in \cite{tetley2019constraining}. Notably, by interpolated hotspot data, we were able to improve the optimization process, allowing for a more direct calibration of the model output to the hotspot trail data. These procedural enhancements not only improved the model's accuracy but also fortify its statistical integrity, thus paving the way for a more objective and reliable understanding of plate movements. 

OptAPM offers a promising route to improved understanding of plate motion. Our central point is that careful formulation of the cost functions and the combined objective function is critical for optAPM and similar optimization approaches to achieve their full potential.

\vspace{3mm}
\noindent {\bf Acknowledgements.}
JU is supported by NSF grant PHY-2209998. This work was completed as part of the MIT PRIMES program. We thank Sebastian von Hausegger and Laura Schaposnik for useful discussions.



\begin{thebibliography}{99}


\bibitem{demets2010geologically}
C.~DeMets, R.~G.~Gordon, and D.~F. Argus.
\textit{Geologically current plate motions.}
Geophys.~J.~Int.~181(1):1-80, 2010.

\bibitem{demets1990current}
C.~DeMets, R.~Gordon, D.~Argus, and S.~Stein.
\textit{Current plate motions.}
Geophys.~J.~Int.~101(2):425-478, 1990.


\bibitem{kusky2013recognition}
T.~Kusky,  \textit{et al.}
\textit{Recognition of ocean plate stratigraphy in accretionary orogens through Earth history: A record of 3.8 billion years of sea floor spreading, subduction, and accretion.}
Gondwana Research, 24(2):501-547, 2013.


\bibitem{muller1997digital}
R.~D.~M\" uller,  \textit{et al.}
\textit{Digital isochrons of the world's ocean floor.}
J.~Geophys.~Res.~Solid Earth, 102(B2):3211, 1997.

\bibitem{Torsvik}
T.~Torsvik, \textit{et al.} \textit{Global plate motion frames: toward a unified model.} Reviews of geophysics 46.3 (2008).


\bibitem{Clennett}
E.~Clennett,  \textit{et al.} \textit{Assessing plate reconstruction models using plate driving force consistency tests.} Scientific reports 13.1 (2023): 10191.


\bibitem{steinberger2008absolute}
B.~Steinberger and T.~Torsvik.
\textit{Absolute plate motions and true polar wander in the absence of hotspot tracks.}
Nature, 452(7187):620-623, 2008.

\bibitem{tetley2019constraining}
M.~G.~Tetley,  \textit{et al.}
\textit{Constraining absolute plate motions since the Triassic.}
J.~Geophys.~Res.~Solid Earth, 124(7):7231-7258, 2019.


\bibitem{Gripp}
A.~Gripp and R.~Gordon. \textit{Young tracks of hotspots and current plate velocities,} Geophys.\ J.\ Int.~150.2 (2002)~321.


\bibitem{doubrovine2012absolute}
P.~Doubrovine, B.~Steinberger, and T.~Torsvik.
\textit{Absolute plate motions in a reference frame defined by moving hot spots in the Pacific, Atlantic, and Indian oceans.}
J.~Geophys.~Res.~Solid Earth, 117(B9), 2012.

\bibitem{Argus}
D.~Argus and R.~Gordon. \textit{No-net-rotation model of current plate velocities incorporating plate motion model NUVEL-1.} Geophys.~Res.~Lett.~18.11 (1991): 2039-2042.

\bibitem{Argus2}
D.~Argus and R.~Gordon, and C.~DeMets. \textit{Geologically current motion of 56 plates relative to the no-net-rotation reference frame.} Geochemistry, Geophysics, Geosystems 12.11 (2011).


\bibitem{gurnis2012plate}
M.~Gurnis,  \textit{et al.}
\textit{Plate tectonic reconstructions with continuously closing plates.}
Computers \& Geosciences, 38(1):35-42, 2012.


\bibitem{seton2012global}
M.~Seton, \textit{et al.}
\textit{Global continental and ocean basin reconstructions since 200 Ma.}
Earth-Science Reviews, 113(3-4):212-270, 2012.


\bibitem{Besse}
J.~Besse and V.~Courtillot. \textit{Apparent and true polar wander and the geometry of the geomagnetic field over the last 200 Myr.} J.~Geophys.~Res.~Solid Earth 107.B11 (2002): EPM-6.


\bibitem{Torsvik-palaeo}
T.~Torsvik, \textit{et al.} \textit{Phanerozoic polar wander, palaeogeography and dynamics.} Earth-Science Reviews 114.3-4 (2012): 325-368.




\bibitem{mitchell2012supercontinent}
R.~Mitchell, T.~Kilian, and D.~Evans.
\textit{Supercontinent cycles and the calculation of absolute palaeolongitude in deep time.}
Nature, 482(7384):208-211, 2012.



\bibitem{TuzoWilson}
J.~T.~Wilson; A Possible Origin of the Hawaiian Islands. Canadian Journal of Earth Sciences 2014; 51 (3): ix-xvi. 

\bibitem{Morgan1}
W.~J.~Morgan, \textit{Convection plumes in the lower mantle.} Nature 230.5288 (1971): 42-43.


\bibitem{Duncan} R.~Duncan, and M.~Richards. \textit{Hotspots, mantle plumes, flood basalts, and true polar wander.} Reviews of Geophysics 29.1 (1991): 31-50.




\bibitem{maher2015absolute}
S.~M.~Maher,  \textit{et al.}
\textit{Absolute plate motion of Africa around Hawaii-Emperor bend time.}
Geophys.~J.~Int.~201(3):1743-1764, 2015.

\bibitem{Steinberger}
B.~Steinberger and R.~J.~O'Connell. \textit{Advection of plumes in mantle flow: implications for hotspot motion, mantle viscosity and plume distribution.} Geophysical Journal International 132.2 (1998): 412-434.


\bibitem{o2005uncertainties}
C.~O'Neill, D.~M\" uller, and B.~Steinberger.
\textit{On the uncertainties in hot spot reconstructions and the significance of moving hot spot reference frames.}
Geochem.~Geophys.~Geosyst.~6(4), 2005.




\bibitem{o2013constraints}
J.~M.~O'Connor,  \textit{et al.}
\textit{Constraints on past plate and mantle motion from new ages for the Hawaiian-Emperor seamount chain.}
Geochem.~Geophys.~Geosyst.~14(10):4564-4584, 2013.


\bibitem{wessel1998geometric}
P.~Wessel and L.~Kroenke.
\textit{The geometric relationship between hot spots and seamounts: Implications for Pacific hot spots.}
Earth Planet.~Sci.~Lett.~158(1-2):1-18, 1998.


\bibitem{knesel2008rapid}
K.~Knesel,  \textit{et al.}
\textit{Rapid change in drift of the Australian plate records collision with Ontong Java plateau.}
Nature, 454(7205):754-757, 2008.

\bibitem{koppers2011new}
A.~Koppers,  \textit{et al.}
\textit{New 40Ar/39Ar age progression for the Louisville hot spot trail and implications for inter-hot spot motion,}
Geochem.~Geophys.~Geosyst.~12(12), 2011.

\bibitem{mcdougall1988age}
I.~McDougall and R.~A.~Duncan.
\textit{Age progressive volcanism in the Tasmantid seamounts.}
Earth and Planetary Science Letters, 89(2):207-220, 1988.


\bibitem{Heuret1}
A.~Heuret and S.~Lallemand. \textit{Plate motions, slab dynamics and back-arc deformation.} Physics of the Earth and Planetary Interiors 149.1-2 (2005): 31-51.


\bibitem{Heuret2}
A.~Heuret, \textit{et al.} \textit{Plate kinematics, slab shape and back-arc stress: A comparison between laboratory models and current subduction zones.} EPSL
256.3-4 (2007): 473-483.


\bibitem{SCHELLART2008118}
W.~P.~Schellart, D.~R.~Stegman, and J.~Freeman.
\textit{Global trench migration velocities and slab migration induced upper mantle volume fluxes: Constraints to find an Earth reference frame based on minimizing viscous dissipation.}
Earth-Science Reviews, 88(1):118-144, 2008.


\bibitem{Lallemand}
S.~Lallemand, \textit{et al.} \textit{Subduction dynamics as revealed by trench migration.} Tectonics 27.3 (2008).

\bibitem{Funiciello}
F.~Funiciello, \textit{et al.} \textit{Trench migration, net rotation and slab-mantle coupling.} Earth and Planetary Science Letters 271.1-4 (2008): 233-240.

\bibitem{Schellart}
W.~Schellart, \textit{A subduction zone reference frame based on slab geometry and subduction partitioning of plate motion and trench migration,} Geophys.~Res.~Lett.~38.16 (2011).



\bibitem{flament2017origin}
N.~Flament,  \textit{et al.}
\textit{Origin and evolution of the deep thermochemical structure beneath Eurasia.}
Nature Communications, 8(1):14164, 2017.


\bibitem{becker2017superweak}
T.~W.~Becker.
\textit{Superweak asthenosphere in light of upper mantle seismic anisotropy.}
Geochem.~Geophys.~Geosyst.~18(5):1986-2003, 2017.





\bibitem{torsvik2010plate}
T.~Torsvik, B.~Steinberger, M.~Gurnis, and C.~Gaina.
\textit{Plate tectonics and net lithosphere rotation over the past 150 My.}
Earth Planet.~Sci.~Lett.~291(1-4):106-112, 2010.


\bibitem{muller2016ocean}
R.~D.~M\" uller,  \textit{et al.}
\textit{Ocean basin evolution and global-scale plate reorganization events since Pangea breakup.}
Annu.~Rev.~Earth Planet.~Sci.~44:107-138, 2016.


\bibitem{Garnero}
E.~Garnero and A.~McNamara. \textit{Structure and dynamics of Earth's lower mantle.} Science 320.5876 (2008): 626.





\bibitem{van2010towards}
D.~G.~Van Der~Meer,  \textit{et al.}
\textit{Towards absolute plate motions constrained by lower-mantle slab remnants.}
Nature Geoscience, 3(1):36-40, 2010.

\bibitem{Hellinger}
S.~Hellinger, \textit{The uncertainties of finite rotations in plate tectonics.} J.~Geophys.~Res.~Solid Earth 86.B10 (1981): 9312-9318.


\bibitem{wang2017bounds}
C.~Wang, R.~G.~Gordon, and T.~Zhang.
\textit{Bounds on geologically current rates of motion of groups of hot spots.}
Geophysical Research Letters, 44(12):6048-6056, 2017.



\bibitem{Tarduno}
J.~Tarduno,  \textit{et al.} \textit{The bent Hawaiian-Emperor hotspot track: Inheriting the mantle wind.} Science 324.5923 (2009): 50-53.


\bibitem{Clouard} V.~Clouard and A.~Bonneville. \textit{How many Pacific hotspots are fed by deep-mantle plumes?.} Geology 29.8 (2001): 695-698.




\bibitem{Kirkwood}
B.~Kirkwood,  \textit{et al.} \textit{Statistical tools for estimating and combining finite rotations and their uncertainties.} Geophysical Journal International 137.2 (1999): 408-428.



\bibitem{Jurdy}
D.~Jurdy and M.~Stefanick. \textit{Errors in plate rotations as described by covariance matrices and their combination in reconstructions.} J.~Geophys.~Res.~Solid Earth 92.B7 (1987): 6310-6318.




\bibitem{merdith2017full}
A.~Merdith, \textit{et al.}
\textit{A full-plate global reconstruction of the Neoproterozoic.}
Gondwana Research, 50:84-134, 2017.


\bibitem{GPlates}
D.~M\"uller, \textit{et al.} \textit{GPlates: Building a virtual Earth through deep time.} Geochem.~Geophys.~Geosyst.~19.7 (2018): 2243-2261.

\bibitem{Boyden}
J.~Boyden,  \textit{et al.} \textit{Next-generation plate-tectonic reconstructions using GPlates.} Geoinformatics: Cyberinfrastructure for the solid earth sciences 9 (2011): 5-114.

\bibitem{Muller}
D.~M\"uller, \textit{et al.}
\textit{A tectonic-rules based mantle reference frame since 1 billion years ago-implications for supercontinent cycles and plate-mantle system evolution.} Solid Earth Discussions 2022 (2022): 1-42.

\bibitem{Shimin}
S.~Wang,  \textit{et al.} \textit{Absolute plate motions relative to deep mantle plumes.} Earth Planet.~Sci.~Lett.~490 (2018): 88.


\bibitem{COBYLA}
M.~Powell,  \textit{A view of algorithms for optimization without derivatives.} Mathematics Today-Bulletin of the Institute of Mathematics and its Applications 43.5 (2007): 170.



\bibitem{johnson2021nlopt}
S.~Johnson and J.~Schueller.
\textit{Nlopt: Nonlinear optimization library.}
Astrophysics Source Code Library, pages ascl-2111, 2021.

\bibitem{Conrad}
C.~Conrad and M.~Behn. {\em Constraints on lithosphere net rotation and asthenospheric viscosity from global mantle flow models and seismic anisotropy.} Geochem.~Geophys.~Geosyst.~11.5 (2010).

\bibitem{Becker}
T.~Becker, {\em On the effect of temperature and strain-rate dependent viscosity on global mantle flow, net rotation, and plate-driving forces}, Geophys.J.Int.~167.2 (2006)~943.


 \end{thebibliography}
\end{document}